\documentclass[a4,11pt]{article}

\usepackage{epsfig}
\usepackage{placeins}
\usepackage{array}
\usepackage{amsmath}
\usepackage[utf8]{inputenc}
\usepackage{multirow}
\usepackage[table]{xcolor}
\usepackage{mdwlist}
\usepackage{natbib}

\begin{document}

\begin{center}
{\Large
	{\sc
	 Bayesian Variable Selection for Probit Mixed Models Applied to Gene Selection
	}
}
\bigskip

Meïli Baragatti$^{1,2,*}$

\medskip
{\it
 $^1$ Ipsogen SA, Luminy Biotech Entreprises, Case 923, Campus de Luminy, 13288 Marseille Cedex 9, France.\\
 $^2$ Institut de Mathématiques de Luminy (IML), CNRS Marseille, case 907, Campus de Luminy, 13288 Marseille Cedex 9, France.\\
 $^*$ baragatt@iml.univ-mrs.fr, baragattimeili@hotmail.com.
}
\end{center}

\begin{center}
PREPRINT
\end{center}

\bigskip
\noindent

\begin{abstract}
	In computational biology, gene expression datasets are characterized by very few individual samples compared to a large number of measurements per sample. Thus, it is appealing to merge these datasets in order to increase the number of observations and diversify the data, allowing a more reliable selection of genes relevant to the biological problem. Besides, the increased size of a merged dataset facilitates its re-splitting into training and validation sets. This necessitates the introduction of the dataset as a random effect. In this context, extending a work of \citet{LeeSha}, a method is proposed to select relevant variables among tens of thousands in a probit mixed regression model, considered as part of a larger hierarchical Bayesian model. Latent variables are used to identify subsets of selected variables and the grouping (or blocking) technique of \citet{Liu} is combined with a Metropolis-within-Gibbs algorithm \citep{MonteCarloStatMethods}. The method is applied to a merged dataset made of three individual gene expression datasets, in which tens of thousands of measurements are available for each of several hundred human breast cancer samples. Even for this large dataset comprised of around 20000 predictors, the method is shown to be efficient and feasible. As an illustration, it is used to select the most important genes that characterize the estrogen receptor status of patients with breast cancer.
\end{abstract}
{\it Keywords}: Bayesian variable selection, random effects, probit mixed regression model, grouping technique (or blocking technique), Metropolis-within-Gibbs algorithm.

\section{Introduction}

Selection of variables is a common problem in many scientific fields, and particularly in bioinformatics. Gene expression profiling analyses are notorious for generating a very large number of predictors compared to the number of observations. Microarray or high throughput sequencing technologies are important for finding genes that are implicated in biological processes including development, disease, and response to treatment, and it plays an important role in the current tendency towards personalized medicine. Identified genes or sequences can be used to classify future observations, influencing the treatment of patients. However, these experiments are expensive, and datasets have often no more than 100 specimens. The goal, therefore, is to advance a method allowing variable selection from merged microarray datasets, each of them presenting its own individual experimental bias.\\

Several model-based approaches have been developed to select variables. A well-known example is SVM (Support Vector Machine) with a recursive feature elimination of the genes (\citet{GuyonWeston}). \citet{GeorgeMcCulloch} and \citet{ChipmanGeorge} developed  Bayesian variable selection with the use of Gibbs sampling for linear models; a review of this type of selection is provided by \citet{OHara}. \citet{TadesseSha2005} proposed a Bayesian variable selection in a model-based clustering approach, using a multivariate Gaussian mixture model. Recently \citet{BottoloRichardson} proposed an algorithm based upon Evolutionary Monte Carlo.
Binary responses are often encountered in biostatistics studies, therefore probit or logistic models are implied. Bayesian variable selection methods have been proposed by \citet{LeeSha}, \citet{ShaVannucci}, \citet{ZhouWang1}, \citet{ZhouWang2} and \citet{YangSong} for probit regression, and by \citet{ZhouLiu},  \citet{ChenDey} and \citet{Tuchler} for logistic regression. Extension to multi-category data has been done for the probit model in \citet{AlbertChib}.\\

The motivation behind the variable selection method developed in this paper is to take the design of the study into account by using random effects in a mixed model. It is particularly suited to a merged microarray dataset design, and many such datasets are freely available from the NCBI GEO website \citep{GEO}. The increased size of a merged dataset may provide improved power, and facilitates its re-splitting into training and validation sets. In addition a merged set comprises more data diversity than an individual set, hence we can avoid bias due to a particular dataset as explained by various authors, see \citet{MetaAnalysis} and references therein.
Among all the methods previously proposed for variable selection, that of \citet{Tuchler} considered mixed models. However, her approach was specific for logistic models, and the method was applied to datasets with only few dozens predictors, whereas the aim of this paper is to select a few predictors among tens of thousands in a Bayesian framework. Recently \citet{FruhwirthWagner} considered variable selection for random effects, but in this paper we are more interested by variable selection for the fixed effects, assuming that random effects are present.
\\

The approach developed in this paper extends the approach of \citet{GeorgeMcCulloch} and \citet{LeeSha}. \citet{GeorgeMcCulloch} introduced latent variables to identify subsets of selected variables in a linear model. Then \citet{LeeSha} used these latent variables in a probit regression model, which is considered as part of a larger hierarchical Bayesian model. Our method extends the model used by \citet{LeeSha} by adding random effects. We are then confronted with several difficulties. One concerns the simulation of conditional distributions, since full conditional distributions cannot be directly simulated. A solution is to use the grouping (or blocking) technique of \citet{Liu}, and to combine Gibbs sampler and Metropolis-Hastings algorithms. Therefore the algorithm developed is a combination of the grouping method of Liu and the Metropolis-within-Gibbs algorithm \citep{MonteCarloStatMethods}. A computational difficulty due to the large number of genes had also been overcome by imposing a fixed number of selected genes at each iteration of the algorithm. As a consequence the influence of the value chosen for \textit{the variable selection coefficient} of our model is reduced. That represents an advantage, since the value of this coefficient can impact the results of other methods, see for instance \citet{BottoloRichardson} who proposed to put a hyperprior distribution on this coefficient.\\

In this paper, Affymetrix microarray data are used, so predictors (genes) will be referred to as ``probesets'', according to that technology.
An Affymetrix U133plus2 microarray profiles all of the genes in the human genome, many of them more than once, using over 54000 gene-specific ``probesets''.
Our Bayesian variable selection method for probit mixed models is developed to select a few important probesets, among tens of thousands, which are indicative of the activity of the estrogen receptor gene in breast cancer. The severity of this common and deadly disease is directly related to estrogen receptor (ER) status, which is traditionally measured biochemically.\\
Three different breast cancer datasets were used, all with clinically defined ER status. One microarray experiment was done per patient, and ten of thousands of probesets were measured per experiment.
The dataset is introduced as a random effect in the model, thus accounting for the different experimental conditions implicit in each set. The three merged datasets were split into training and validation sets, and the relevance of the selected probesets was checked by fitting a probit mixed model on the training set and predicting the ER status for the patients from the validation set and other independent sets available from the NCBI GEO website. The stability and the sensitivity of the algorithm were also checked by using the relative weighted consistency measure of \citet{Somol2008}.\\

The remainder of the paper is organized as follows. Section 2 describes the probit mixed model with latent variables. Section 3 gives the full conditional distributions necessary for the Gibbs sampling algorithm, outlines the algorithm and proposes a way to construct a classification rule using the selected probesets. Section 4 provides some experimental results on real datasets, on the relevance of selected probesets, and on the sensitivity and the stability of the method. Finally Section 5 discusses the method.

\section{Probit mixed model for gene selection}
\subsection{The hierarchical model}
	Suppose that $n$ binary events are observed, denoted by the $Y_i$, $i = 1,\ldots,n$.
	The set of potential regressors is of size $p$, with $p \gg n$. The goal is to select a subset of regressors related to the events $Y_1,\ldots,Y_n$. The following probit mixed model is considered,
	\begin{displaymath}
		P(Y_i=1  \mid  U, \beta)=p_i=\Phi(X_i'\beta + Z_i'U),
	\end{displaymath}
	where $\Phi$ stands for the standard Gaussian cumulative distribution function, and $X_i$ and $Z_i$ for the fixed and random effect regressors associated with the $i^{th}$ observation. The parameter $\beta$ corresponds to the fixed-effect coefficients and the parameter $U$ to the random-effect coefficients. $X$ and $Z$ are design matrices associated with the fixed and random effects.\\
	Assuming that we have $K$ random effects, $U=(U_1',\ldots,U_K')'$. Each $U_l$ is of size $q_l$, and $\sum_{l=1}^K q_l = q$. The size of $\beta$ is $p$.
	\vspace{0.4cm}
	
	Following \citet{AlbertChib} and \citet{LeeSha}, a vector of latent variables $L$ is introduced. We write $L=(L_1,\ldots,L_n)$ and we assume that $L  \mid  U, \beta \sim \mathcal{N}_n(X\beta+ZU, I_n)$ with $I_n$ the identity matrix. We then have
	\begin{equation}\label{LatentVar}
		Y_i = \left\{
		\begin{array}{rl}
			1 & \text{if } L_i>0\\
			0 & \text{if } L_i<0,
		\end{array} \right.
	\end{equation} 
% 	giving
% 	\begin{displaymath}
% 		P(Y_i=1  \mid  U, \beta)=P(L_i>0  \mid  U, \beta)=\Phi(X_i'\beta+Z_i'U)=p_i.
% 	\end{displaymath}
% 	\vspace{0.2cm}
	
	To perform variable selection, a vector $\gamma$ of $p$ indicator variables is introduced:
	\begin{equation*}
		\gamma_j = \left\{
		\begin{array}{rl}
			1 & \text{if } \beta_j \neq 0, \quad \text{variable }j \text{ selected} \\
			0 & \text{if } \beta_j = 0, \quad \text{variable }j \text{ not selected}.
		\end{array} \right.
	\end{equation*} 
	Given $\gamma$, $\beta_{\gamma}$ is the vector of all nonzero elements of $\beta$, and $X_{\gamma}$ is the matrix $X$ with only the columns corresponding to the elements of $\gamma$ that are equal to 1.

\subsection{Prior distributions}	
	To complete the hierarchical model, some prior assumptions have to be made on $U \mid D$, $\beta_{\gamma} \mid \gamma$, $\gamma$ and $D$, where $D$ is a covariance matrix of dimension $q$.
	\begin{itemize}
		\item If the data supports $\gamma_j = 0$ over $\gamma_j = 1$, then the $j^{th}$ variable will not be needed in the model and we can let  $\beta_j=0$. We then focus on the prior distribution of the non null vector $\beta_{\gamma}$. Like \citet{LeeSha}, we take the following conventional prior:
			\begin{equation}\label{priorbetagamma}
				\beta_{\gamma} \mid \gamma \sim \mathcal{N}_d(0,c(X_{\gamma}'X_{\gamma})^{-1}), \qquad \textrm{with} \qquad d=\sum_{j=1}^p \gamma_j,
			\end{equation}
			This prior corresponds to the g-prior of \citet{Zellner86}, and $c$ is a positive scale factor specified by the user. \citet{BottoloRichardson} called it the variable selection coefficient. Several authors discussed the choice of its value, see \citet{ChipmanGeorge}, \citet{GeorgeFoster}, \citet{ClydeGeorge} and \citet{SmithKohn} among others. %It assumed that the prior distribution is as if it resulted from a prior experiment with the same model matrix $X$. reflecting similar correlation structures. This prior reflect the covariance structure of the likelihood. 
			\citet{Raftery97} used a similar form of prior. In our algorithm the value of $c$ will be fixed, but will not be too influent (see the discussion).

		\item The $\gamma_j$ are assumed to be independent Bernoulli variables, with 
			\begin{displaymath}
				P(\gamma_j=1)=\pi_j, \qquad 0 \leq \pi_j \leq 1.
			\end{displaymath}
			We do not want to use prior knowledge to favor any probesets, so we put $\quad \pi_j=\pi$, $\forall j=1,\ldots p$.

		\item The vector of coefficients associated with the random effects is assumed to be Gaussian and centered:
			\begin{displaymath}
				U \mid D \sim \mathcal{N}_{q}(0,D).
			\end{displaymath}

			This definition allows three cases to be distinguished:\\

			\emph{General case:} No structure is assumed for the variance-covariance matrix $D$, its prior distribution is an Inverse-Wishart $\mathcal{W}^{-1}(\Psi,m)$.\\

			\emph{Case of a block-diagonal matrix $D$:} The different random effects are assumed independent. The vectors of coefficients associated with each random effect have Gaussian prior distributions:
			\begin{displaymath}
				U_l \mid A_l \sim \mathcal{N}_{q_l}(0,A_l), \quad l = 1,\ldots,K,
			\end{displaymath}
			where the $A_l$ are symmetric design matrices of dimension $q_l$. $D$ is a block-diagonal matrix denoted by $diag(A_1,\ldots,A_K)$. The prior distributions for each $A_l$ are Inverse-Wishart $\mathcal{W}^{-1}(\Psi,m)$.\\

			\emph{Case of a diagonal matrix $D$:} $D=diag(A_1,\ldots,A_K)$ where $A_l=\sigma_l^2 I_{q_l}$, $l=1,\ldots,K$ and $I_{q_l}$ the identity matrix. The prior distributions for the $\sigma_l^2$ are then Inverse Gamma $\mathcal{IG}(a,b)$ ($b$ denotes the scale).
	\end{itemize}

\section{Bayesian sampler for variable selection}
\subsection{The conditional distributions}
	The posterior distribution of $\gamma$ is of particular interest since it encapsulates the effectiveness of the different explanatory variables in explaining the variation in the responses $Y$. The number of possible explanatory variables is on the order of tens of thousands, so the number of possible $\gamma$-vectors is extremely large. The idea is to use a Gibbs sampling algorithm to explore this posterior distribution and search for high probability $\gamma$ values.\\
	%The simulated Markov Chain for $\gamma$ would have the property that high probability $\gamma$ values appear more frequently than low probability values, thereby facilitating the search for more "promising" models. Relevant variables can be quickly identified while the sampler explores the posterior distribution.
	% See George et McCulloch1993 p884, et Chipma, George, McCulloch 2001 p90.
	%It follows that the length of the Markov chain can be much smaller than $2^p$ and still serve to identify at least some of the high probability values. \\

	In order to use the classical Gibbs sampler, we must be able to simulate from all of the full conditional distributions (simplified by the hierarchical structure): $f(L \mid Y,\beta,U)$, $f(\beta \mid L,U,\gamma)$, $f(U \mid L,\beta,D)$, $f(\gamma \mid L,U,\beta)$ and $f(D \mid U)$.
	
	%The sampler explores the posterior distribution. One can quite quickly identify useful variables which have high marginal probabilities of $\gamma_j=1$.

	\begin{itemize}
	\item Full conditional distribution of $L$.
%		Combining $L_i  \mid  U, \beta \sim \mathcal{N}(X_i'\beta+Z_i'U,1)$ with (\ref{LatentVar}), we obtain
		\begin{eqnarray}\label{fullL}
		 \nonumber L_i \mid \beta,U,Y_i=1 &\sim& \mathcal{N}(X_i'\beta+Z_i'U,1) \quad \textrm{left truncated at 0}\\
		 L_i \mid \beta,U,Y_i=0 &\sim& \mathcal{N}(X_i'\beta+Z_i'U,1) \quad \textrm{right truncated at 0}.
		\end{eqnarray}

	\item Full conditional distribution of $\beta$.\\
		Given $\gamma$, we know which elements of $\beta$ are not null. So we focus on the generation of the non null elements of $\beta_{\gamma}$. Letting $V_{\gamma}=\frac{c}{1+c}(X_{\gamma}'X_{\gamma})^{-1}$, we have
		\begin{eqnarray}\label{fullbeta}
		 	\beta_{\gamma} \mid L,U,\gamma &\sim& \mathcal{N}_d(V_{\gamma}X_{\gamma}'(L-ZU), V_{\gamma})	 \qquad \textrm{with} \qquad d=\sum_{i=1}^p \gamma_i.
		\end{eqnarray}

	\item Full conditional distribution of $U$.\\
		Defining $W=(Z'Z+D^{-1})^{-1}$, we have
		\begin{eqnarray}\label{fullU}
		 	U \mid L,\beta,D \sim \mathcal{N}_q(WZ'(L-X\beta), W).
		\end{eqnarray}

	\item Full conditional distributions of $\gamma$.
		%We know the distribution of $\beta_{\gamma} \mid \gamma$, but not of $\beta \mid \gamma$. Using $f(\gamma \mid \beta_{\gamma},L,U)$ instead of $f(\gamma \mid \beta,L,U)$, we obtain
		\begin{eqnarray}\label{fullgamma}
 			\nonumber f(\gamma \mid \beta_{\gamma},L,U)  &\propto& (2\pi)^{-\frac d2} \exp\Big[-\frac12 \Big(-L'X_{\gamma}\beta_{\gamma}-\beta_{\gamma}'X_{\gamma}'L+\beta_{\gamma}'X_{\gamma}'ZU+U'Z'X_{\gamma}\beta_{\gamma}+\beta_{\gamma}' V_{\gamma}^{-1} \beta_{\gamma}\Big)\Big]\\
			& & \times  \mid c(X_{\gamma}'X_{\gamma})^{-1} \mid ^{-\frac 12} \prod_{j=1}^p \pi_j^{\gamma_j} (1-\pi_j)^{1-\gamma_j}.
		\end{eqnarray}

	\item Full conditional distribution of $D$.\\
		\emph{General case:} The full conditional distribution of $D$ is an Inverse-Wishart:
		\begin{eqnarray}\label{fullD1}
			D \mid U  &\sim& \mathcal{W}^{-1}(UU'+\Psi,m+1).
		\end{eqnarray}

		\emph{Case of a block-diagonal matrix $D$:} $D=diag(A_1,\ldots,A_K)$. The full conditional distribution of $A_l$ ($\forall l=1,\ldots,K$) is an Inverse-Wishart:
		\begin{eqnarray}\label{fullD2}
			A_l \mid U_l  &\sim& \mathcal{W}^{-1}(U_lU_l'+\Psi,m+1).
		\end{eqnarray}

		\emph{Case of a diagonal matrix $D$:} $D=diag(A_1,\ldots,A_K)$, and $\forall l=1,\ldots,K$, $A_l=\sigma_l^2 I_{q_l}$. The full conditional distribution of $\sigma_l^2$ is an Inverse-Gamma:
		\begin{eqnarray}\label{fullsigma}
			\sigma_l^2 \mid U_l  &\sim& \mathcal{IG}\Big(\frac{q_l}{2}+a,\big(\frac 12 U_l'U_l+b\big)\Big).
		\end{eqnarray}
	\end{itemize}

\subsection{Use of the grouping technique}
	The classical Gibbs sampler cannot be used because the full conditional distribution of $\gamma$ cannot be directly simulated (see (\ref{fullgamma})). However, this full conditional distribution can be simulated with a Metropolis-Hastings algorithm, and the complete algorithm would be a Metropolis-within-Gibbs algorithm. \citet{RobertsRosenthal2006} have shown the Harris-recurrence of this algorithm, therefore its convergence is guaranteed.  But even with a Metropolis-Hastings algorithm, the full conditional distribution of $\gamma$ is difficult to obtain, since it depends on the actual value of $\beta_{\gamma}$. Thus the acceptance rate for a candidate $\gamma^*$ in the Metropolis-Hastings algorithm will depend both on the actual $\gamma^{(t)}$ and $\beta_{\gamma^{(t)}}$, and on the proposed $\gamma^*$ and $\beta_{\gamma^*}$. The problem is that $\beta_{\gamma^*}$ is unknown.\\
	To get around this problem, we combine the Metropolis-within-Gibbs algorithm with the grouping (or blocking) technique of \citet{Liu}. The idea is to group the parameters $\beta_{\gamma}$ and $\gamma$, so we will be interested in the full conditional distribution of $(\beta_{\gamma},\gamma) \mid L,U$. This technique improves the algorithm and facilitates the convergence of the Markov chain, see \citet{Liu} and \citet{vanDyk}. We note that the sampler obtained is then a special case of a Partial Collapsed Gibbs Sampler, see \citet{vanDyk}.\\
	As we have
	\begin{displaymath}
	f(\beta_{\gamma},\gamma \mid L,U) \propto f(\gamma \mid L,U) f(\beta_{\gamma} \mid \gamma,L,U) ,
	\end{displaymath}
	we remark that simulating from the full conditional distribution $(\beta_{\gamma},\gamma) \mid L,U$ is equivalent to simulating $\gamma$ from its full conditional distribution integrated on $\beta_{\gamma}$, then simulating $\beta_{\gamma}$ from its full conditional distribution. The ``integrated distribution'' for $\gamma$ will not depend anymore on the nuisance parameter $\beta_{\gamma}$ and will be easily simulated by a Metropolis-Hastings algorithm.\\
	In each iteration of the algorithm, we will take care to simulate $\gamma$ before $\beta{\gamma}$, to keep the dependence between $\beta{\gamma}$ and $\gamma$, as noted by \citet{vanDyk}.\\

	We use $f(L \mid \gamma,U)$ and the Bayes Theorem to get the integrated distribution of $\gamma \mid L,U$ (the target distribution):
	\begin{eqnarray}\label{marggamma}
		f(\gamma \mid L,U) &\propto& (1+c)^{-\frac {\sum \gamma_i}{2}} \exp\Big[-\frac12 \Big\{(L-ZU)'(L-ZU) \\
		\nonumber  & & -\frac{c}{1+c}(L-ZU)'X_{\gamma}(X_{\gamma}'X_{\gamma})^{-1}X_{\gamma}'(L-ZU)\Big\}\Big] \times \prod_{j=1}^{p} \pi_j^{\gamma_j}(1-\pi_j)^{1-\gamma_j}.
	\end{eqnarray}

\subsection{The Metropolis-within-Gibbs sampler modified by the grouping technique}
	\subsubsection{A Metropolis-Hastings step to simulate $\gamma$}\label{MHgamma}
		At iteration $(i+1)$ of the Metropolis-Hastings algorithm, a candidate $\gamma^*$ will be proposed from $\gamma^{(i)}$. We want a symmetric transition kernel, to simplify the acceptance rate of the algorithm. The simplest way to have a symmetric transition kernel is to propose a $\gamma^*$ which corresponds to $\gamma^{(i)}$ in which $r$ components have been randomly changed (see \citet{ChipmanGeorge} and \citet{GeorgeMcCulloch97}).\\

		Given the target distribution (\ref{marggamma}), the acceptance rate $\rho$ is then:
		\begin{eqnarray}\label{acceptancerate}
		 \nonumber \rho(\gamma^{(i)},\gamma^*) &=& 
			min\Bigg\{\exp\Big[\frac{c}{2(1+c)}(L-ZU)'\Big(X_{\gamma^*}(X_{\gamma^*}'X_{\gamma^*})^{-1}X_{\gamma^*}'-X_{\gamma^{(i)}}(X_{\gamma^{(i)}}'X_{\gamma^{(i)}})^{-1}X_{\gamma^{(i)}}' \Big) \\
			& & \times (L-ZU)\Big] \times (1+c)^{\frac{\sum \big(\gamma_j^{(i)}-\gamma_j^*\big)}{2}} \times \Big(\frac{\pi}{1-\pi}\Big)^{\sum_1^p \big(\gamma_j^*-\gamma_j^{(i)}\big)},1\Bigg\}.
		\end{eqnarray}

		To facilitate the computation of the algorithm, the proposed $\gamma^*$ still corresponds to $\gamma^{(i)}$ for which $r$ components have been changed, but in such a way that the number of components whose values are 1 (and so the number of selected variables) is invariant. In so doing, $r/2$ components among the 1 values, and $r/2$ components among the 0 values are chosen at random and switched. There are several advantages to propose such a $\gamma^*$:
		\begin{itemize}
		\item[$\bullet$] In an iteration of the algorithm, if we have the number of variables selected $d$ higher than the number of observations $n$, the $X_{\gamma}'X_{\gamma}$ matrix would be singular, and the prior distribution of $\beta_{\gamma}$ could not be defined as in (\ref{priorbetagamma}). An advantage of fixing the number of variables to be selected at each iteration is that this number cannot increase during a run of the algorithm, and if $d$ is chosen lower than $n$ this case of non singularity of $X_{\gamma}'X_{\gamma}$ is avoided.
		\item[$\bullet$] The acceptance rate is simplified, as we obtain $\sum \big(\gamma_j^{(t)}-\gamma_j^*\big)=0$.
		\item[$\bullet$] The choice of the prior value for the variable selection coefficient $c$ used in the prior distribution of $\beta$ is less influent (see the discussion).
		\end{itemize}

		\textbf{Remark.} In the method of \citet{LeeSha}, the $\gamma$ vector is generated component by component at each iteration, while in our method a Metropolis-Hastings algorithm is used to generate it. There are two advantages to use a Metropolis-Hastings algorithm: it is computationally advantageous for a very large number of variables compared to a generation component by component, and it enables us to easily generate a $\gamma$ vector with an invariant number of components whose values are 1.

	\subsubsection{The sampler}
	The Metropolis-within-Gibbs sampler modified by the grouping technique of Liu generates a sequence:
	\begin{displaymath}
	 \gamma^{(1)},\beta_{\gamma}^{(1)},D^{(1)},L^{(1)},U^{(1)},\ldots \ldots, \gamma^{(b+m)},\beta_{\gamma}^{(b+m)},D^{(b+m)},L^{(b+m)},U^{(b+m)}.
	\end{displaymath}
% 	\begin{displaymath}
% 	 \gamma^{(1)},\beta_{\gamma}^{(1)},D^{(1)},L^{(1)},U^{(1)},\ldots, \gamma^{(t)},\beta_{\gamma}^{(t)},D^{(t)},L^{(t)},U^{(t)},\ldots, \gamma^{(b+m)},\beta_{\gamma}^{(b+m)},D^{(b+m)},L^{(b+m)},U^{(b+m)}.
% 	\end{displaymath}
	The sequence of the $\gamma^{(t)}$, which is of interest for the variable selection problem, is embedded in this "Gibbs sequence". To generate it, at each iteration $\gamma$ is simulated from its integrated distribution and $\beta_{\gamma},L,U$ and $D$ are simulated from their full conditional distributions.
	\begin{quote}
	 \textbf{Algorithm:}\\
	Starting with initial values $\gamma^{(0)},\beta^{(0)},D^{(0)},L^{(0)},U^{(0)}$.
	At iteration $t+1$:
		\begin{enumerate}
		 \item Simulate $\gamma^{(t+1)}$ from $f(\gamma \mid L^{(t)},U^{(t)})$ (see \ref{marggamma}), using the Metropolis-Hasting (MH) step. The MH step begins with $\gamma^{(t)}$ as an initial value, and $k$ iterations are performed given $L^{(t)}$ and $U^{(t)}$ ($k$ arbitrarily fixed). At iteration $i+1$ of the MH step:
			\begin{enumerate}
			\item Generate the $\gamma^*$ candidate, by randomly switching $r/2$ components among the 1 values, and $r/2$ components among the 0 values.
			\item Take \begin{equation*}
					\gamma^{(i+1)} = \left\{
					\begin{array}{rl}
					\gamma^* & \text{with probability} \qquad \rho(\gamma^{(i)},\gamma^*) \qquad  \text{see (\ref{acceptancerate})}\\
					\gamma^{(i)} & \text{with probability} \qquad 1-\rho(\gamma^{(i)},\gamma^*)
					\end{array} \right.
				\end{equation*}
			\end{enumerate}
			$\gamma^{(t+1)}$ will be the $\gamma^{(k)}$ obtained at the $k^{th}$ iteration of the MH step.
		\item Simulate $\beta_{\gamma}^{(t+1)}$ from $f(\beta_{\gamma} \mid L^{(t)},U^{(t)},\gamma^{(t+1)})$ (see (\ref{fullbeta})).
		\item Simulate $D^{(t+1)}$ from $f(D \mid U^{(t)})$ (see (\ref{fullD1}), (\ref{fullD2}) or (\ref{fullsigma})).
		\item Simulate $L^{(t+1)}$ from $f(L \mid Y,\beta^{(t+1)},U^{(t)})$ (see (\ref{fullL})).
		\item $U^{(t+1)}$ from $f(U \mid L^{(t+1)},\beta^{(t+1)},D^{(t+1)})$ (see (\ref{fullU})).
		\end{enumerate}
	\end{quote}
	We use the fact that $X\beta=X_{\gamma}\beta_{\gamma}$ and that $\beta$ can be obtained from $\gamma$ and $\beta_{\gamma}$.
	The number of iterations is $b+m$, where $b$ corresponds to the burn-in period and $m$ to the observations from the posterior distributions.

	In our application, we are not concerned by the strict convergence of the sampler. The aim is to find some relevant variables explaining the response, and to obtain good predictions. Hence we only need to do stability and sensitivity studies to check that the training set and the choices of the hyperparameters are not too influent, and we check the biological relevance of the variables selected.

	\subsubsection{The selected probesets}
	For selection of variables, the sequence $\{\gamma^{(t)}=(\gamma_1^{(t)},\ldots,\gamma_p^{(t)}),t=b+1,\ldots,b+m\}$ is used. The most relevant variables for the regression model are those which are supported by the data and prior information. Thus they are those corresponding to the $\gamma$ components with higher posterior probabilities%. In addition, the empirical frequencies of the $\gamma$ values are consistent estimates of their probabilities according to the posterior distribution. Therefore the relevant variables appear most frequently in the simulated observations from the posterior distribution of $\gamma$
	, and can be identified as the $\gamma$ components that are most often equal to 1.

\subsection{Classification and prediction}
	Once a set of relevant variables have been selected, it can be used to fit a probit mixed model in a classical way and to classify future observations. However, if more variables than necessary to fit a probit mixed model have been selected in the Bayesian selection step, a second selection has to be performed on them in order to build a reliable probit mixed model. This second selection is performed on the training set using standard selection tools like AIC, BIC, Bayes factors,\ldots. The final probit mixed model can be tested on the validation set. Moreover, the variables selected in the Bayesian selection step can be used in other classification methods, such as Support Vector Machines (but random effects are not taken into account).

\section{Experimental results}
\subsection{Application to the ER status of patients with breast cancer}
	\subsubsection{Description of the datasets}
		Three different datasets were used: one private dataset from the Institut Paoli Calmettes (Marseille, France), consisting of 151 samples, and two datasets freely available from the NCBI GEO public website \citep{GEO}: accession numbers GSE2109 (310 samples) and  GSE5460 (124 samples). Each dataset was treated for background noise and normalized with respect to a reference distribution by the RMA procedure \citep{Irizarry}. Each dataset was split into a training set and a validation set having the same proportions of ER positive and ER negative observations. Then the three training datasets were merged on one side (497 patients) and the three validation sets were merged on the other side (88 patients). %A variable was created to represent the random effect corresponding to datasets.
		\\
% 		In clinical practice, ER status is measured biochemically and reclassed as positive or negative in order to make treatment decisions. However, the parameters used to define this binary reclassification may depend on the medical center and on the country. For consistency this clinical classification cannot be used. The expression of the ESR1 probeset (Affymetrix symbol \texttt{205225\_at}) was then used for this study: patients with high expression of this gene were considered positive (>9), and patients with low expression were considered negative (<7.7). These cut-offs were determined after cross-checking with the clinical ER status.\\
		For each patient, more than 54000 probesets were available. Two filters were applied on all of these probesets. Only the probesets sufficiently expressed so that they can be differentiated from noise and which could not be considered as invariant were kept, resulting in 19384 probesets. The goal was to select only a few probesets which are related to the ER status of the patients, by taking into account the different experimental conditions between the different merged datasets.
%		Selected probesets should correspond to genes in biochemical pathways that are known to be effected by ER expression.
		\vspace{0.4cm}

		In this illustration, there are thousands of fixed regressors corresponding to the expression measurements of probesets, and only one random effect, which corresponds to the different datasets. $X_{ij}$ corresponds to the measurement of the expression level of the $j^{th}$ probeset for the $i^{th}$ patient, and $Z_{il}=1$ if the $i^{th}$ patient is from the $l^{th}$ dataset, 0 otherwise.

	\subsubsection{Prior settings for the algorithm}
		\begin{itemize}
		\item Following the recommendations of \citet{SmithKohn}, a value of $c=50$ was chosen for the variable selection coefficient used in the prior distribution of $\beta$.
		\item Thirty probesets were selected at each iteration of the Gibbs sampler, when $\gamma$ is generated; $r=10$ of them were changed at each iteration of the Metropolis-Hastings step (5 zeros and 5 ones).
		\item The random effect corresponds to the dataset, and the three datasets are considered independent: they were generated in different countries, by different teams, using different equipments and different patients. Therefore the variance-covariance matrix of the random effect $D$ was a diagonal matrix $3 \times 3$ with $A_1=\sigma_1^2 I_{3}$. \citet{Gelman2006} noted that an inverse-gamma prior should not be too non-informative, otherwise serious problems can arise. Given our data, we knew that $\sigma_1^2$ is probably not too high, and a $\mathcal{IG}(2,3)$ seemed reasonable for the prior distribution of $\sigma_1^2$.
		\item For the Metropolis-within-Gibbs sampler modified by the grouping technique, 60000 iterations were computed, among which 30000 were burn-in iterations. For the Metropolis-Hastings step in this sampler, 500 iterations were computed, and the simulated $\gamma$ was the one corresponding to the $500^{th}$ iteration.
		\end{itemize}

	\subsubsection{Results and predictions}\label{ResultsPredictions}
\begin{figure}[!h]
			\begin{center}
			\includegraphics[scale=0.4]{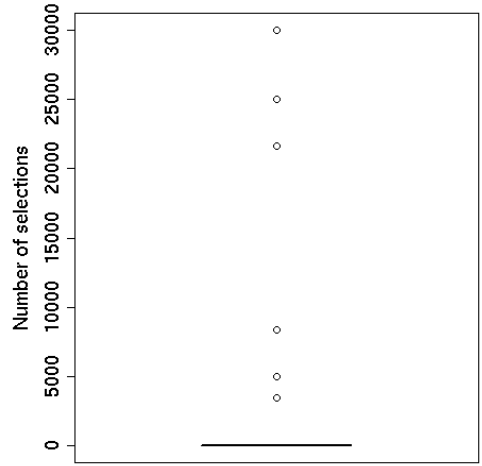}
			\end{center}
			\caption{Boxplot of the number of selections of a probeset after the burn-in period, for the real datasets example. Forty probesets were selected at least once, all of the other probesets were never selected. A point represents a probeset (or several probesets if they have been selected the same number of times).}
			\label{Boxplot:boxplotrealdata}
		\end{figure}

		We performed a first selection of variables by selecting the top-rank probesets, those which have been selected the most often. A boxplot can help, see Figure \ref{Boxplot:boxplotrealdata}. Forty probesets were selected at least once from the 30000 post-burn-in iterations of the simulated Markov Chain for $\gamma$. Twenty three probesets were selected in the 30000 iterations, and thirty were selected at least from 20000 iterations. There is a gap between probesets selected in more than 20000 iterations and others, so the first selection is made of these probesets selected at least in 20000 iterations.\\
		A second selection was performed on the thirty probesets from the first selection, to build a reliable probit mixed model. This second selection was performed on the training set, using a stepwise selection (with AIC and BIC criteria) and the classification performance of the model on the validation set. Five probesets were kept: Affymetrix symbols \texttt{228241\_at}, \texttt{205862\_at}, \texttt{202376\_at}, \texttt{216222\_s\_at} and \texttt{1568760\_at}. See Table \ref{Tab:tab1} for the associated gene symbols and coefficients. The estimated random effects of this final model were reasonable: $-0.284$ for the dataset from the Institut Paoli Calmettes, $0.199$ for the GSE2109 dataset and $0.087$ for the GSE5460 dataset. 
		%The random part was small compared to the fixed part of the model, indicating that the merged dataset was not too heterogeneous, and that the merging was not inappropriate.

		\begin{table}[h!]
		\begin{center}
		\begin{tabular}{|c|c|c|c|}
		\hline
		Probeset & gene & Coefficient & Pvalue\\
		\hline
		Intercept & & -9.12074 &   1.92e-05\\
		\texttt{228241\_at} & AGR3 &  0.45046  &  1.12e-15\\
		\texttt{205862\_at} & GREB1 &  0.77639  &  4.18e-08\\
		\texttt{202376\_at} & SERPINA3 &   0.37965  &  0.000149\\
		\texttt{216222\_s\_at} & MYO10 & -0.63551 &   0.004967\\
		\texttt{1568760\_at} & MYH11 &  0.42742  &  0.050219\\
		\hline
		\end{tabular}
		\caption{Probesets selected in the final model and associated coefficients.}
		\label{Tab:tab1}
		\end{center}
		\end{table}
		\FloatBarrier

\begin{figure}[!h]
			\begin{center}
			\includegraphics[scale=0.42]{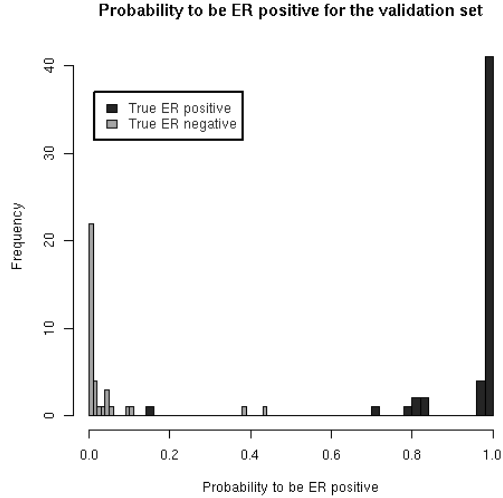}
			\end{center}
			\caption{Histogram of probabilities to be ER positive given by the final model, for patients from the validation set.}
			\label{Fig:histoPosNegEn}
		\end{figure}

		Using this 5-probeset model, two methods were used to predict the ER status of the patients in the validation set:
		\begin{enumerate*}
		\item Using knowledge of the dataset to which each patient belonged and using the estimated random effects coefficients.
		\item The estimated random effects coefficients are not used in order to mimic a real-life scenario of an experiment for a patient coming from an unknown dataset.
		\end{enumerate*}
		The patients were predicted positive if their probability to be positive was higher than 0.5 and negative if it was lower than 0.5. The two methods gave us the same predictions, which were very good: a specificity of 1 and a sensitivity of 0.98 (1 wrong predictions among 88),  see Figure \ref{Fig:histoPosNegEn}.\\

		\FloatBarrier
		
		\textbf{Remark.} In biomedical studies, when continuous variables are often reclassified as binary, it is common to define an ``undetermined zone'' of probabilities for which no prediction are given. Indeed, it is sometimes better than giving a wrong prediction, because these predictions imply treatments. Defining an ``undetermined zone'' between 10\% and 90\% probability of being positive, false predictions were eliminated, and 10 were considered undetermined (11.4\%) (estimated random effects coefficients not used).\\

		As a final test of our model, two more independent datasets were brought in from the NCBI GEO website: the GEO series GSE6532 and the GEO series GSE12763. The random effects associated with these datasets were entirely unknown, simulating an even more realistic case of prediction for a patient coming from an unknown dataset. Once again the results were very good : only 1 wrong prediction among 29 for the GSE12763 dataset, and no wrong predictions among 86 for the GSE6532 dataset.

\subsection{Sensitivity and stability studies}
	The sensitivity and the stability of the algorithm were assessed by using the relative weighted consistency measure of \citet{Somol2008}, denoted by $CW_{rel}$. %It would have been possible to use the stability index of \citet{Kuncheva2007}, but in order to use it the subsets of variables selected by the different runs cannot be of varying sizes, which could be the case if reliable selections are wanted. 
	It is a measure evaluating how much subsets of selected variables for several runs overlap, and it shows the relative amount of randomness inherent in the concrete variable selection process. It takes values between 0 and 1, where 0 represents the outcome of completely random occurrence
	of variables in the selected subsets and 1 indicates the most stable variable selection outcome possible.\\
	Stability is defined as sensitivity to variations in the training set. Referring to our breast cancer data set, 4000 probesets were randomly chosen from among the 19384 originally available. Since the aim here was only to check the sensitivity and stability of the method, these 4000 were not chosen in relation to the ER status.\\
	Several runs of the algorithm were performed, and are reported in Table \ref{Tab:tab2}. Concerning the stability, the algorithm was run on three different training sets of 497 microarrays (among 585), using the same prior values for the hyperparameters.
% 	$c=50$. The prior for $\sigma_1^2$ was an $\mathcal{IG}(2,3)$. Fifteen probesets were selected at each iteration of the algorithm. Six were randomly changed at each iteration of the Metropolis-Hastings step, and 12000 iterations were computed. Among the 12000, 6000 were burn-in iterations and 500 were computed for each Metropolis-Hastings step. Four probesets were kept in two runs among three, with stability index $\mathcal{S}=0.266$.
	Concerning the sensitivity, the algorithm was run on the same training set with different values of $c$, different prior distributions for $\sigma_1^2$, different numbers of probesets to be selected at each iteration of the algorithm and different numbers of iterations. For the prior distributions for $\sigma_1^2$, we chose a $\mathcal{IG}(2,3)$ which seemed reasonable given our data, a $\mathcal{IG}(2,5)$ to have a prior favoring higher values compared to the first one, a $\mathcal{IG}(3,1)$ to favor lower values, and a $\mathcal{IG}(1,1)$ to have a non-informative prior without too small parameters to avoid problems, see \citet{Gelman2006}.\\
% 	\begin{itemize}
% 	\item[$\bullet$] Different values of $c$ (rows 4-7 of Table \ref{Tab:tab2}: $c=10$, 50, 100, 1000). Five probesets were kept in at least 2 runs among 4, and $S=0.332$. The probeset 215552\_s\_at was kept in all 4 of the runs.
% 	\item[$\bullet$] Different prior distributions for $\sigma_1^2$ (rows 8-10 of Table \ref{Tab:tab2}: $\sigma_1^2 \sim \mathcal{IG}(2,3)$, $\mathcal{IG}(2,5)$, $\mathcal{IG}(1,1)$). Four probesets were kept in 2 runs among 3, and $S=0.266$.
% 	\item[$\bullet$] Different numbers of probesets to be selected at each iteration of the algorithm (rows 11-13 of Table \ref{Tab:tab2}). These were altered at each iteration of the Metropolis-Hastings step : 15 selected and 6 changed,  5 selected and 2 changed, or 30 selected and 10 changed. Four probesets were kept in at least 2 runs out of 3, and $S=0.399$. Note that these parameters do not seem to modify the number of probesets more selected than the others during the run. The probeset 209603\_at was kept in the three runs.
% 	\item[$\bullet$] Finally, we ran the algorithm with 30000 iterations, among which 15000 were burn-in iterations, to check if the results were different from those obtained previously using 12000 iterations. The 5 probesets kept with 30000 iterations had already been kept in at least one of the other runs with 12000 iterations. Once again note that the length of the Markov chain does not seem to modify the number of probesets more selected than the others during the run.
% 	\end{itemize}

\begin{figure}[!h]
			\begin{center}
			\includegraphics[scale=0.5]{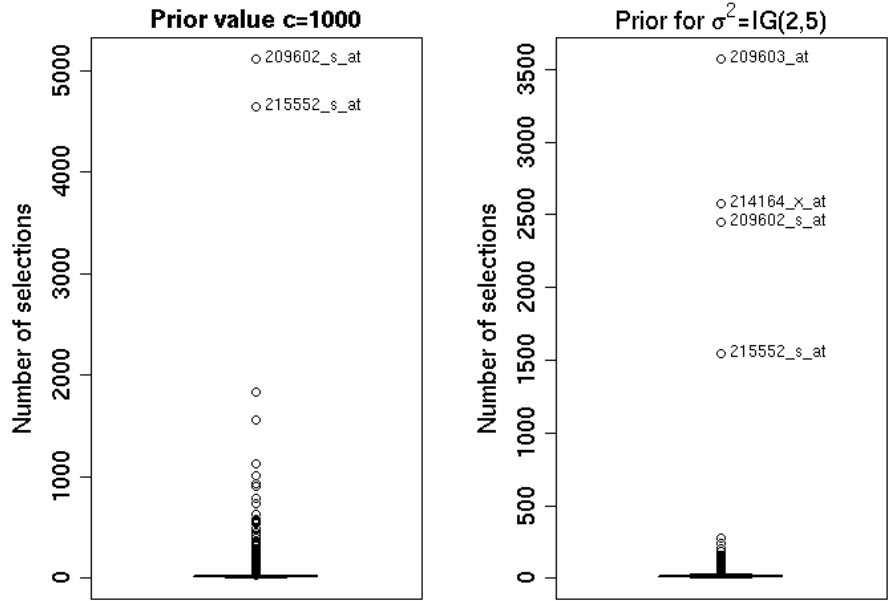}
			\end{center}
			\caption{Boxplot of the number of selections of a probeset after the burn-in period, for two runs of the sensitivity analysis. A point represents a probeset (or several probesets if they have been selected the same number of times). The left boxplot corresponds to the run with $c=1000$: there is a gap between the two probesets selected in more than 4000 iterations and the others, hence we selected these two probesets. The right boxplot corresponds to the run with $\sigma_1^2 \sim \mathcal{IG}(2,5)$: there is a gap between the four probesets selected in more than 1500 iterations and the others, hence we selected these four probesets.}
			\label{Boxplot:boxplotsensitivity}
	\end{figure}

	For each run, a reasonable number of probesets could be easily selected. Indeed, two to ten probesets were selected much more often than the others, see Figure \ref{Boxplot:boxplotsensitivity} (two to four probesets were selected for most of the runs). Hence there is no need to perform a second selection, as in section \ref{ResultsPredictions}. To compare the results of the different runs, the relative weighted consistency measure of Somol and Novovicova $CW_{rel}$ was used.

	\begin{table}[h!]
	\hspace{-2cm}
	\begin{tabular}{|c|c|c|c|c|c|c|c|c|}
	    \hline
	    \rowcolor{lightgray}  &  & Value & Prior & Nb probesets to be  & Nb probesets to be  & Iterations & burn-in &\\
	    \rowcolor{lightgray} Simu & Dataset & of & for & selected at each & changed at each & for the & for the & $CW_{rel}$\\
  	    \rowcolor{lightgray}  &  & $c$ & $\sigma^2$ & iteration of the GS &  iteration of the MH & algo & algo &\\
	    \hline
	    1 & 1 &  &  &  &  &  &  &\\
	    2 & 2 &  &  &  &  &  &  &\\
	    3 & 3 &  \multirow{-3}{*}{50} & \multirow{-3}{*}{$IG(2,3)$} & \multirow{-3}{*}{15} & \multirow{-3}{*}{6} & \multirow{-3}{*}{12000} & \multirow{-3}{*}{6000} & 0.25\\
	    \hline
	    4 &  & 10 &  &  &  &  &  &\\
	    5 &  & 50 &  &  &  &  &  &\\
	    6 &  & 100 &  &  &  &  &  &\\
	    7 & \multirow{-4}{*}{1} &  1000 & \multirow{-4}{*}{$IG(2,3)$} & \multirow{-4}{*}{15} & \multirow{-4}{*}{6} & \multirow{-4}{*}{12000} & \multirow{-4}{*}{6000} & 0.375\\
	    \hline
	    8 &  &  & $IG(2,3)$ &  &  &  &  &\\
	    9 &  &  & $IG(1,1)$ &  &  &  &  &\\
	    10 &  &  & $IG(1,1)$ &  &  &  &  &\\
	    11 & \multirow{-4}{*}{1} & \multirow{-4}{*}{50} & $IG(2,5)$ & \multirow{-4}{*}{15} & \multirow{-4}{*}{6} & \multirow{-4}{*}{12000} & \multirow{-4}{*}{6000} & 0.292\\
	    \hline
	    12 &  &  &  & 15 & 6 &  &  &\\
	    13 &  &  &  & 5 & 2 &  &  &\\
	    14 & \multirow{-3}{*}{1} & \multirow{-3}{*}{50} & \multirow{-3}{*}{$IG(2,3)$} & 30 & 10 & \multirow{-3}{*}{12000} & \multirow{-3}{*}{6000} & 0.5\\
	    \hline
	    15 & 1 & 50 & $IG(2,3)$ & 15 & 6 & 30000 & 15000 & \\
	    \hline
	\end{tabular}
	\caption{Parameters of the runs for the stability and sensitivity study and associated relative weighted consistency measure of Somol and Novovicova $CW_{rel}$. For the Metropolis-Hastings step, always 500 iterations are computed.}
	\label{Tab:tab2}
	\end{table}
	\FloatBarrier

	Using the results of the 15 runs together, $CW_{rel}=0.398$. Subsets of selected variables for the different runs overlapped:  among the 15 runs the probesets \texttt{215552\_s\_at}, \texttt{209603\_at} and \texttt{209602\_s\_at} were kept in 12, 12 and 6 runs respectively. Apparently the prior for $\sigma^2$ ($CW_{rel}=0.292$) has more impact than the number of probesets to be selected at each iteration ($CW_{rel}=0.5$). We note that the number of probesets to be selected at each iteration of the algorithm and the number of iterations do not seem to modify the number of probesets more selected than the others during the run.\\

	This was satisfying and the method appears relatively stable. First because the random selection of 4000 probesets carries a risk of destabilization of the results, since these 4000 are not necessarily those which are most indicative of ER status. Secondly, several probesets can represent the same gene, and different genes can be implied in the same biological pathway. Thus, it is possible that subsets of probesets are more similar than they appear, and therefore that $CW_{rel}$ is underestimated. For example, the probesets \texttt{209603\_at} and \texttt{209602\_s\_at} mentioned above both represent the gene GATA3. Finally, these simulations indicate that there is not a parameter whose choice introduces more sensitivity than the others.\\

\subsection{Comparison with other methods}
	We compared the performance of our method with the performances of other methods which do not take into account random effects: we considered the model of \citet{LeeSha} and Support Vector Machine with Recursive Feature Elimination of the variables, with linear or non linear kernels \citep{GuyonWeston}. We used simulated data: 200 observations of 1000 variables following a uniform distribution $\mathcal{U}_{[-5,5]}$ are generated. We assume that 5 variables and a random effect $U$ of size 4 explain a vector of binary variables $Y$ by a probit mixed model:
	\begin{eqnarray*}
	 p_i &=& \Phi(X_i'\beta + Z_i'U), \qquad i=1,\ldots,200\\
	 Y_i &\sim& \mathcal{B}(p_i), \qquad i=1,\ldots,200
	\end{eqnarray*}
	We took $\beta_{\gamma}=(-1,-1,1,1,2)$ and we assume that 50 observations are coming from each modality of the random effect. Different values of $U$ were used: $U1=(0,0,0,0)$, $U2=(-3,-2,2,3)$, $U3=(-5,-3,3,5)$, $U4=(-10,-5,5,10)$ and $U5=(-30,-10,10,30)$. This set of 200 observations was splitted into training and validation sets, each of them of size 100, with 25 observations coming from each modality of the random effect. For our method and the method of Lee et al. we took $c=50$, 5 probesets were selected at each iteration of the Gibbs sampler and $r=2$ of them were changed at each iteration of the Metropolis-Hastings step (1 zero and 1 one), $D$ was a diagonal matrix $3 \times 3$ with $A_1=\sigma_1^2 I_{3}$ and a prior  $\mathcal{IG}(1,1)$ was chosen for $\sigma_1^2$, 500 iterations were performed for each Metropolis-Hastings step, and a total of 3000 and 5000 iterations were performed for the whole algorithm.\\
 	Concerning our method and the method of \citet{LeeSha}, the top-ranked variables (variables selected more often than others, box-plots were used) were used to perform predictions on the validation set. The RFE-SVM method gave us directly sets of "best variables" and associated models, and these models were used to perform the predictions. The results obtained are in Table \ref{Tab:tab3}.

	\begin{table}[h!]
	\hspace{-2cm}
 	\begin{tabular}{|c|c|c|c|c|c|c|}
 	    \hline
 	    \rowcolor{lightgray} Random effect & \multicolumn{2}{c|}{\cellcolor{lightgray} Our method} & \multicolumn{2}{c|}{\cellcolor{lightgray} \citet{LeeSha} method} & \multicolumn{2}{c|}{\cellcolor{lightgray} RFE-SVM}\\
	    \hline
 	    \rowcolor{lightgray} $U$ & 3000 iterations & 5000 iterations & 3000 iterations & 5000 iterations & linear & non linear\\
 	    \hline
 	    $U1$ & 17 & 26 & 19 & 22 & 25 & 23\\
 	    \hline
 	    $U2$ & 19 & 21 & 19 & 19 & 20 & 26\\
 	    \hline
 	    $U3$ & 21 & 23 & 24 & 24 & 25 & 26\\
 	    \hline
 	    $U4$ & 19 & 19 & 35 & 35 & 29 & 31\\
 	    \hline
 	    $U5$ & 14 & 11 & 44 & 44 & 52 & 56\\
 	    \hline
 	\end{tabular}
	\caption{Number of misclassifications on the validation set, for different methods and different random effects.}
 	\label{Tab:tab3}
 	\end{table}
 	\FloatBarrier

	When there is no random effect or when the magnitude of the random effect is small, our method is comparable to the one of Lee et al., and the results of these two methods are better than or comparable with those obtained by RFE-SVM. But when the magnitude of the random effect is high, especially for $U4$ and $U5$, it appears that our method outperforms the method of Lee et al. and the RFE-SVM method.

\section{Discussion}
	In this article we have developed an approach for Bayesian gene selection for a probit mixed model, as an extension of previous works by \citet{GeorgeMcCulloch} and \citet{LeeSha}. An important contribution of our method is that it allows selection of variables in a mixed framework, taking into account the design of the data. It is particularly useful for  gene selection, as it enables the use of merged datasets in order to introduce more observations and greater diversity. That may provide improved power, and we can avoid bias due to a particular dataset.
	The increased size of a merged dataset facilitates its re-splitting into training and validation sets, hence we do not need to evaluate the performance of a classification rule by a cross-validation procedure. It is advantageous compared to other methods which do not take into account random effects. Indeed, as these methods can use only one dataset which is usually of small size, they often need to perform leave-one-out-cross-validation, which can be time-consuming (see \citet{LeeSha}, \citet{YangSong}, \citet{ShaVannucci}, \citet{ZhouWang1} and \citet{ZhouWang2} for instance). On the contrary, if several datasets are merged then a separated training set can be used and the performance of a classifier can be directly obtained on it.
	Using simulations to make comparisons with other methods which do not take into account random effects, we showed that the proposed method is comparable to others when the magnitude of the random effects is low, but performs better than the others for classification when the magnitude of the random effects is high. This method should prove widely useful in microarray bioinformatics, since many diverse datasets are freely available on the Internet. But it can also be used for data obtained from high throughput sequencing technologies, which will probably be used a lot in few years. Indeed, the method can be applied when we have a matrix with $n << p$ and an associated vector of random effects.\\

	In practice, before running an analysis, one must decide how many variables will be selected at each iteration of the Gibbs sampler. We do not consider this to be a drawback, since in order to have a reliable selection, the number of probesets should be limited compared to the size of the training set. Besides, fixing the number of variables selected at each iteration is a computational advantage, as discussed in section \ref{MHgamma}. In particular, the singularity of the $X_{\gamma}'X_{\gamma}$ matrix is avoided.\\
%	The number of variables to be changed at each iteration of the Metropolis-Hastings step, $r$, is also chosen as a prior value. This value must be large enough to allow thorough mixing of the vector $\gamma$. Fortunately, the sensitivity study showed that the algorithm is not overly sensitive to the choice of the number of variables selected at each iteration of the Gibbs sampler and to the choice of $r$.\\
% 	While 60000 iterations may appear small compared to 19384 probesets, the vector of interest $\gamma$ is well mixed between two iterations, and all the more so when $r$ is large. Indeed, inside an iteration of the Gibbs sampler, several iterations of the Metropolis-Hastings algorithm are computed, and inside each a new $\gamma$ vector of $r$ components is proposed. As a consequence, the $\gamma$ vectors between two iterations of the Gibbs sampler were fairly different during the burn-in period. Furthermore, we were not concerned by the strict convergence of the sampler, as the aim was to find relevant variables and to obtain good predictions. Therefore, we just checked with the sensitivity study that the results obtained for a selection among 4000 probesets were similar for 12000 and 30000 iterations.\\
	 In addition, one must choose a value for the hyperparameter $c$ which is large enough to have a relatively non-informative prior. Only the simulations of $\beta_{\gamma} \mid L, U, \gamma$ and of $\gamma \mid L,U$ directly depend on $c$. Concerning $\beta_{\gamma} \mid L, U, \gamma$, we can see in (\ref{fullbeta}) that the density is proportional to $c/(1+c)$, which is relatively close to 1 if $c$ is large. Concerning the density of $\gamma \mid L,U$ (see (\ref{marggamma})), it depends on $c/(1+c)$ and on $(1+c)^{-\frac{\sum \gamma_i}{2}}$. The factor $(1+c)^{-\frac{\sum \gamma_i}{2}}$ does not play a role in the simulation of $\gamma$, because the number of variables to be selected at each iteration is fixed: this factor vanishes in the acceptance rate of the Metropolis-Hastings step of the algorithm. Therefore the value chosen should not be too influent, as long as it is large enough.
	 We chose arbitrarily $c=50$, following Smith and Kohn's (1997) recommendations. However different authors suggested different ranges, see \citet{ChipmanGeorge}, \citet{GeorgeFoster} and \citet{ClydeGeorge} among others. For example \citet{ZhouWang1} and \citet{ZhouWang2} used $c=10$, and \citet{LeeSha} used $c=100$. However, it is possible to include another level in our Bayesian hierarchical model and to put a prior distribution on $c$. \citet{ZellnerSiow} for instance proposed a mixture of g-priors and an inverse-gamma prior on $c$. Recently \citet{BottoloRichardson} considered putting a hyperprior on $c$ and using a Metropolis-within-Gibbs with adaptive proposal for updating this coefficient.
	 In our application this coefficient was held fixed for convenience, and good results were obtained. Besides the sensitivity study showed us that the method is not overly sensitive to the value chosen for $c$, as expected.\\
	 More generally, it appeared that the algorithm is fairly stable to variations in the training set, and is robust to prior value of any of the hyperparameters.\\

	Convergence could not be verified because we did not have formal diagnostic tools to prove it, as the parameters vectors used in the proposed algorithm were not associated to the same variables from one iteration to the next. Besides, the different runs could have converged to a local mode of the posterior distribution of $\gamma$, and not to a global one. But the results obtained in the stability and sensitivity analyses were satisfactory, as different runs with different starting points and different prior hyperparameters selected broadly the same variables, which means that these different chains had basically the same behavior. From our experience, it appeared that having a total number of iterations equal to three times the size of the set of predictors is sufficient, the results were not significantly different when more iterations were performed.\\

	The probesets selected by our method to characterize the estrogen receptor status enabled us to fit a model with good predictions. Moreover, three genes among the five used in the model were also selected using a Support Vector Machine method (twenty-four genes were selected by SVM), and another group of three among those five is known to be associated with estrogen receptor pathways and breast cancer: GREB1 \citep{Nagaraja, Towson, Rae}, SERPINA3 \citep{Cimino} and MYH11 \citep{Singh}. Therefore, it seems that the probesets selected by our method are quite biologically relevant.\\

	The algorithm developed is efficient and feasible, even for very large datasets with around 20000 variables. Therefore this approach has a clear advantage over other selection methods which handle less variables or which do not take into account random effects. However, Bayesian variable selection is an active research area, and it would be interesting to combine our method with recent proposals. For instance by studying the performance of the method with other prior distributions for $\sigma^2$, like half-Cauchy or folded-noncentral-\textit{t} distributions, see \citet{Gelman2006}. Or by putting a prior distribution on $c$, like in \citet{BottoloRichardson}. It would also be of interest to consider an alternative prior distribution for $\beta_{\gamma}$ to handle a non-invertible $X_{\gamma}'X_{\gamma}$ (when $\gamma$ is itself singular or when $n<d$), by combining our approach with the concept of ridge regression (work in progress,  \citet{BaragattiPommeretRidge}).

 	\section{Acknowledgements}
 	We would like to thank an Associate Editor and two anonymous reviewers for careful reading of the paper and constructive comments which have led to an improvement of the manuscript.
 	We are grateful to Dr Daniel Birnbaum and Pr François Bertucci from the Département d'Oncologie Moléculaire of the Institut PAOLI CALMETTES (Marseille, France) for permission to use their data.
 	We also thank Pr Denys Pommeret and Rebecca Tagett for useful discussions and comments.

\bibliography{references}

\end{document}